\documentclass{ws-ijmpcs}

\def\draftnote{}

\usepackage{graphicx}
\usepackage{url}
\usepackage[super]{cite}

\makeatletter
\def\openaccesstext{}
\makeatother

\begin{document}
	
	\markboth{Z. Roupas}{Black hole spin-mass correlation at birth stage of star clusters}
		
	\title{BLACK HOLE SPIN-MASS CORRELATION AND ACCRETION DISKS AT BIRTH STAGE OF DENSE STAR CLUSTERS}
	
	\author{ZACHARIAS ROUPAS}
	
	\address{Dipartimento di Fisica ``G. Occhialini'', 
		Universit\'a degli Studi di Milano-Bicocca, \\
		Piazza della Scienza 3, 20126 Milano, Italy\\
		zacharias.roupas@unimib.it}
		
	\address{Istituto Nazionale di Fisica Nucleare (INFN), Sezione di Milano-Bicocca, \\
		Piazza della Scienza 3, 20126 Milano, Italy}
	
	\maketitle
		
	\begin{abstract}
		We review the accretion-driven channel of stellar black-hole growth and spin-up in gaseous star clusters, developed within the framework of our MSCA project ``protoBH''. Our detailed calculations involving semi-analytic modelling, dynamical and hydrodynamic simulations suggest that the growth and spin-up of initially low-mass stellar black holes is an integral element of the formation process of dense massive star clusters, such as the proto-globular clusters observed at high redshift. The black holes accrete residual gas, remaining from the first star formation event, and form accretion disks generated by gas angular momentum induced by the black-holes' own curved orbits. Assuming initial black hole masses below the pair-instability mass gap $\leq 55\,{\rm M}_{\odot}$, the gap gets populated with black-hole masses growing to $\sim 10^2\,{\rm M}_{\odot}$, and even reaching several hundred solar masses for sufficiently dense and massive proto-clusters. We predict a consistent spin-mass relation with low black hole spins $\lesssim 0.2$ for low black hole masses $\lesssim 20\,{\rm M}_{\odot}$, spanning all intermediate spins for higher masses and saturating to high spins $\gtrsim 0.8$ for high masses $\gtrsim 70\,{\rm M}_{\odot}$. We find evidence in favour of such an accretion-driven subpopulation of black holes in the GWTC-5.0 released by the LIGO-Virgo-KAGRA collaboration.
	\end{abstract}
		
	\section{Introduction}
	
	The astrophysical origin of heavy black holes (BHs) $\sim 10^2-10^3\,{\rm M}_{\odot}$ is an open, active area of research \cite{2019A&A...632L...8R,2020PhRvD.102d3002B,2020ApJ...904L..13R,Kimball_2020,2020ApJ...903L..21S,2021MNRAS.505..339M,2022MNRAS.512..884R,2023MNRAS.526..429A,2024A&A...688A.148T,2024Sci...384.1488F,2024AJ....167..191P,2025PhRvD.111f3039B,2025ApJ...988...15L,2025ApJ...994L..37K,2025ApJ...993L..54G,2025ApJ...993L..30L,2025A&A...702A.208R,2026A&A...709A.5R,2026ApJ...996L..44B,2026arXiv260717869R}. The theorized upper BH mass gap induced by the physics of a pair-instability supernova (PISN)\cite{2016A&A...594A..97B,2017MNRAS.470.4739S,2021ApJ...912L..31W} excludes BHs within the mass range $\sim 60-130\,{\rm M}_{\odot}$ generated directly by stellar evolution. Nevertheless, the gravitational-wave (GW) experiments LIGO-Virgo-KAGRA (LVK) have revealed a significant population of BHs within the mass gap\cite{2026arXiv260527225T}. Several attempts have been made to identify BH subpopulations in gravitational-wave transient catalogues (GWTC)\cite{2022PhRvD.105l3024F,2025PhRvL.134a1401A,2025ApJ...994..261A,2026Natur.652..874T,2026arXiv260509351B,2026arXiv260524281H,2026arXiv260527226T,2026arXiv260623305R,2026PhRvL.137b1403B,2026PhRvL.137b1404P,2026ApJ..1005L..55R,2026arXiv260600234P,2026ApJ..1007L..45W,2026ApJ..1001L..40F,2026arXiv260829072R}, related to astrophysical mechanisms that can generate heavy BHs.
	
	Recently, we identified a spin-mass correlation of BHs, generated by accretion in dense, massive star proto-clusters\cite{2026A&A...709A.5R}. 
	Direct observations by the James Webb Space Telescope (JWST) revealed five proto-globular clusters of age $\sim 50\,{\rm Myr}$ in the Cosmic Gems arc galaxy at high redshift $\sim 10$ \cite{2024Natur.632..513A}, which present low metallicity and are remarkably compact. Their typical masses are $\sim 10^6\,{\rm M}_\odot$ and half-light radii $\sim 1\, {\rm pc}$. Our mechanism operates in the early birth stage of such a proto-cluster, at an age $\lesssim 10-20\,{\rm Myr}$, when residual gas from the first star-formation event has not yet been depleted by stellar winds and supernova (SN) explosions. Lower metallicity implies weaker stellar winds, allowing for longer deplettion timescales.
	Stellar BHs, originating from the most massive stars, are born so early in the life of a sufficiently compact star proto-cluster that they have enough time to accrete the residual gas\cite{2025A&A...702A.208R}. Their mass can grow via accretion to values within the mass gap and above. They also spin up\cite{2026A&A...709A.5R}, because as the Bondi sphere of each BH is advected along the BH orbit, angular momentum is injected into the sphere, induced by the velocity shear between the inner and outer hemispheres\cite{2026arXiv260712465R}. 
	This accretion process generates a certain spin-mass relation spanning all BH spin values, from low spins $\lesssim 0.2$ at low masses to saturated high spins $\sim 0.9$ at high masses $\gtrsim 70\,{\rm M}_{\odot}$ with a transition region at masses $\sim 45\,{\rm M}_{\odot}$ with intermediate spin values. We have identified evidence of this accretion-driven channel in the GWTC-5.0 released by LVK\cite{2026arXiv260829072R}.
	
	In the next section we discuss the accretion-disk formation mechanism around stellar BHs in gaseous star clusters\cite{2026arXiv260712465R}. In section \ref{sec:spinmass} we present the BH spin-mass correlation generated by accretion in those clusters\cite{2026A&A...709A.5R}. In section \ref{sec:gwtc} we report the evidence in the GWTC-5.0 for a BH subpopulation generated by this accretion-driven channel\cite{2026arXiv260829072R}. We conclude in the final section.
	
	\section{Accretion-disk formation}\label{sec:disk}
	
	Consider a stellar BH ($m_{\bullet} \sim 10-100\,{\rm M}_{\odot}$) moving in a regular orbit inside the core of a gaseous cluster. Assuming that the gas mass density is nearly constant deep in the core where the BH moves, the BH orbit is a closed elliptical (or circular) curve, generated by a harmonic potential, about the center of the cluster\cite{2026arXiv260712465R}.
	The velocity shear between the inner and outer hemispheres of the Bondi sphere supplies angular momentum to the gas as the sphere is advected along the curved BH trajectory. This gas angular momentum generates an accretion disk around the BH.
	
\begin{figure}[tbp]
		\centering
		\includegraphics[width=0.32\textwidth]{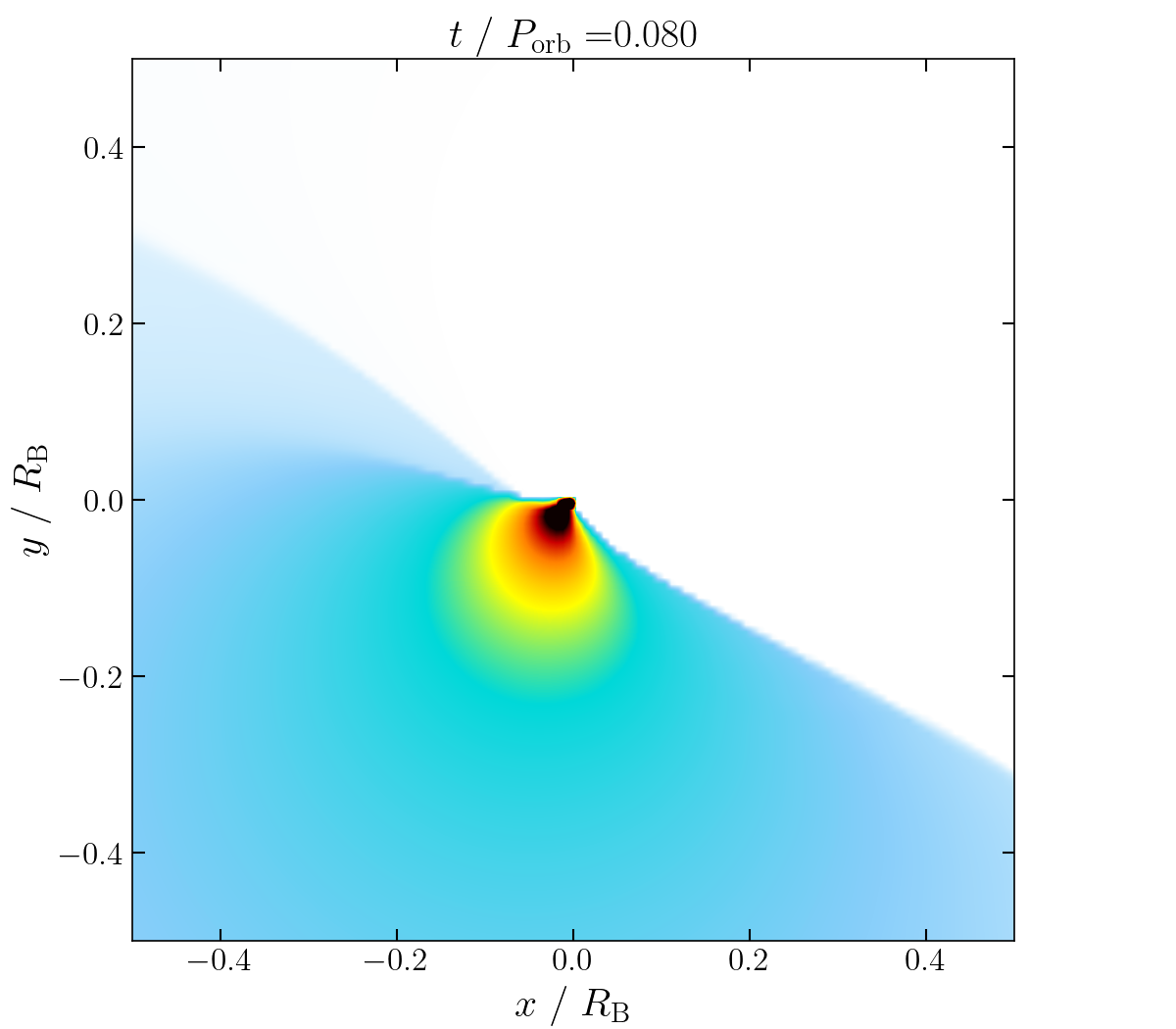}
		\includegraphics[width=0.32\textwidth]{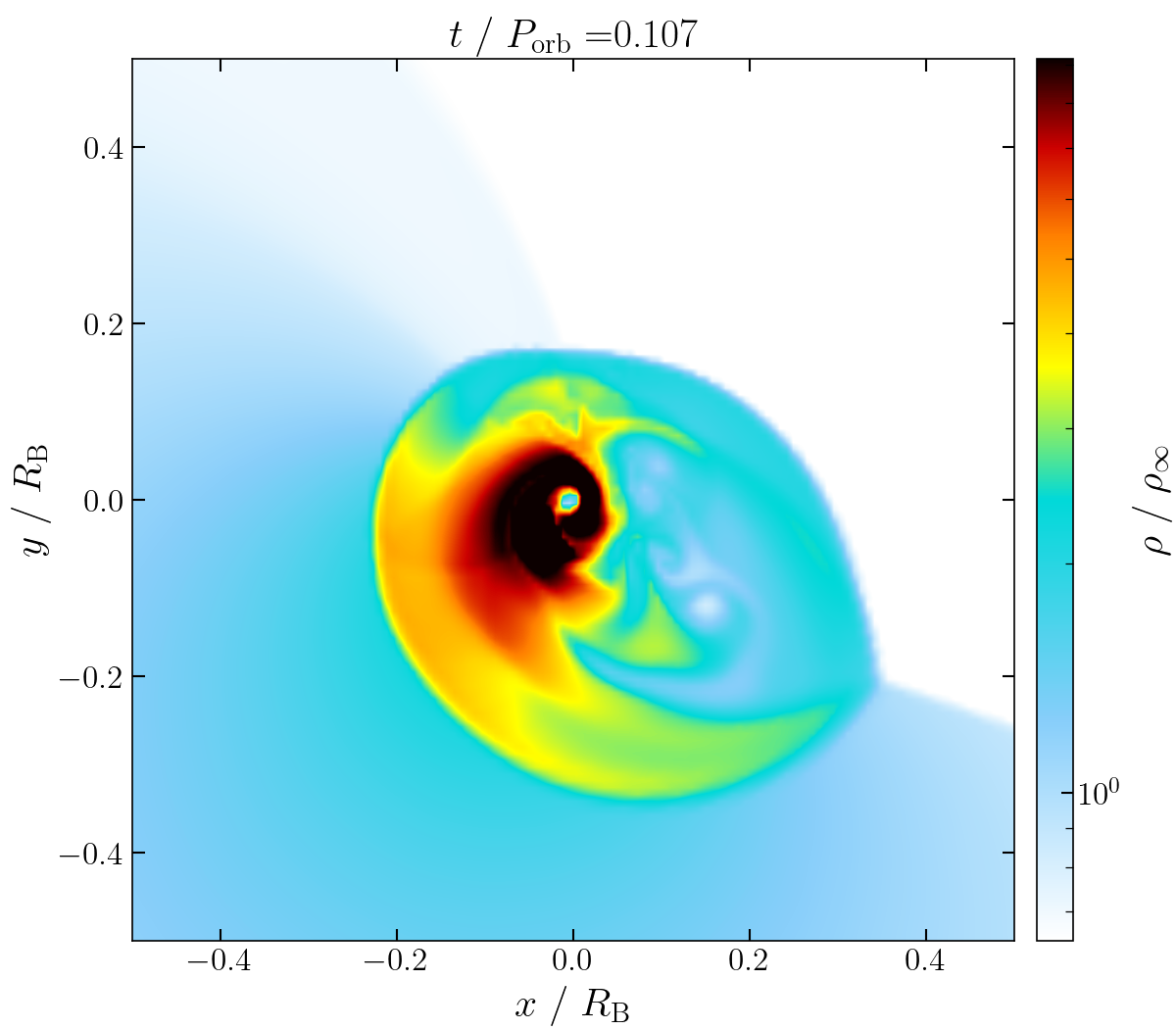}
		\includegraphics[width=0.32\textwidth]{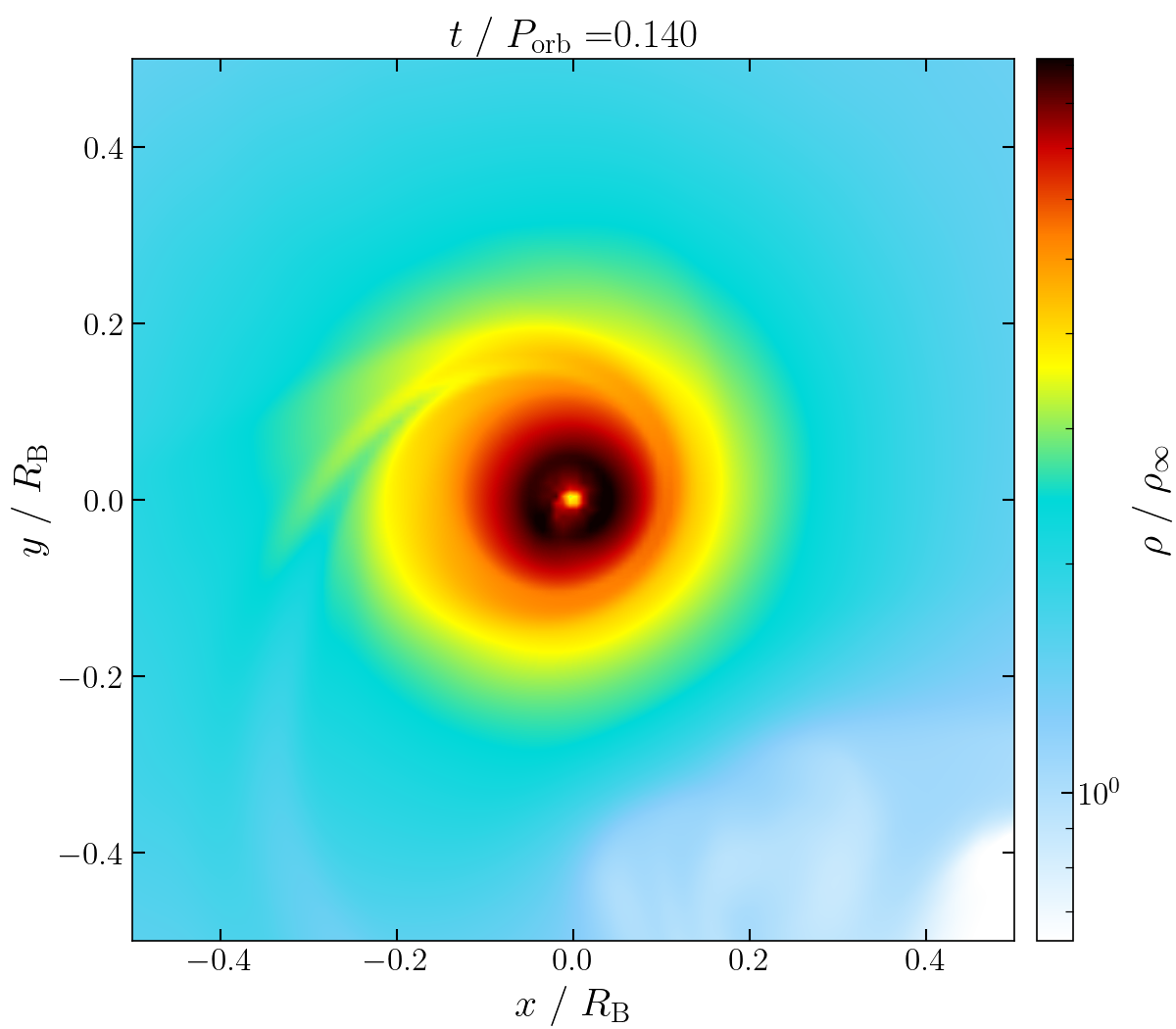}	
		\caption{The gas density contour at different times, in units of the orbital period $P_{\rm orb}$, for a 2D simulation run with BH mass $m_{\bullet} = 50{\rm M}_{\odot}$, cluster density $\rho_{\rm cl} = 10^7{\rm M}_{\odot}/{\rm pc}^3$, and a circular BH orbit. The BH orbital angular momentum vector is directed out of the page towards the reader (cluster center is at the left side of each panel). We observe a density front with a rotation tendency counter to the BH orbital rotation. The steady state of a disk, counter-rotating to the BH orbit, is reached in the right panel.}
		\label{fig:movie_2D}
	\end{figure}

	We verified this picture with 2D and 3D hydrodynamic numerical simulations\cite{2026arXiv260712465R}, using \textsc{Athena++} v21.0\cite{2020ApJS..249....4S,athena_v21}\footnote{\url{https://github.com/PrincetonUniversity/athena}}, as depicted in Fig.~\ref{fig:movie_2D}. Starting from an initially uniform, static gas configuration we find that an accretion disk is formed fast within a fraction of the BH orbital period\cite{2026arXiv260712465R}, 
	$\tau_{\rm d} \sim 0.1 P_{\rm orb}$, for realistic compact dense clusters with mass $M_{\rm cl} \sim 10^{5-7}\,\rm{M}_{\odot}$, core size $r_c\sim 0.1-10.0 \,{\rm pc}$, and star formation efficiency $\varepsilon \sim 0.1-0.3$. The disk radius is of the same order as the circularization radius 
	\begin{equation}
		R_{\rm d} \sim R_{\rm circ}\equiv \frac{\Omega^2 R_{\rm B}^4}{G m_{\bullet}},
	\end{equation}	
	where $\Omega = \sqrt{4\pi G \rho_{\rm cl}/3}$ is the BH orbital angular velocity determined by the cluster density $\rho_{\rm cl}$, and $R_{\rm B}$ the Bondi radius defining the BH gravitational sphere of influence. This suggests that $R_{\rm d} \sim (10^{-4}-10^{-2}) R_{\rm B}$ for the clusters of interest and BHs moving inside the core. The accretion disks around the BHs are slim with disk aspect ratio $H/R \sim 0.2-0.4$. See Ref.~[\refcite{2026arXiv260712465R}] for more details.
			
	\section{Spin-mass correlation}\label{sec:spinmass}
	
	We consider gaseous, dense, massive star clusters at subsolar $\sim 0.1\, {\rm Z}_{\odot}$ and low $\lesssim 0.01\, {\rm Z}_{\odot}$ metallicities modelling the early formation stage $\lesssim 20\,{\rm Myr}$ of massive star clusters, while residual gas from the first star formation event is still present. 
	Proto-globular clusters of age $\sim 50,{\rm Myr}$ have been observed at high redshift with JWST with typical masses $\sim 10^6\,{\rm M}_{\odot}$ and sizes $\sim 1\,{\rm pc}$\cite{2024Natur.632..513A}. 
	Our semi-analytic model and simulation setup follow the BH spin-up and mass growth\cite{2025A&A...702A.208R,2026A&A...709A.5R} from the birth of the first BHs (death of the most massive stars) up to the time when $99\%$ of the residual gas is depleted, which for the assumed cluster compactness is $\sim 13\,{\rm Myr}$. 
	We calculated the depletion timescale for each simulation run explicitly by estimating the stellar winds and supernova energy injection.
	In our runs each BH is created at the time set by the lifetime of its zero-age main sequence progenitor and is assigned the related generation process (direct collapse, core collapse supernova, pair-instability supernova) through the ZAMS-to-BH mass correspondence of Ref.~[\refcite{2017MNRAS.470.4739S}]. The initial BH masses are  drawn from a Salpeter mass function with an upper cut-off at $55\,{\rm M}_{\odot}$. Their initial positions and velocities are sampled from the gravitational Brownian probability density\cite{2021A&A...646A..20R}. 
	The BH orbits are integrated with a drift-kick scheme. The deterministic forces of gravity and dynamical friction are evaluated with a Runge-Kutta method, using the Chandrasekhar formula for the stellar component, its hydrodynamic counterpart for the gas, and the deceleration due to the mass increase. Stochastic velocity increments are applied at adaptive timesteps, modeled as white noise for the stellar encounters and as an Ornstein--Uhlenbeck process for the gas turbulence. Those timesteps, calculated dynamically for each BH throughout the run, have values in effect $>0.5\,P_{\rm orb}$, so that an accretion disk is able to regain a stationary state (see previous section) following a stochastic kick.
	The accretion is of isotropic Bondi--Hoyle type, with the accretion radius taken as the smaller of the Bondi--Hoyle and tidal-lobe radii at each instant, and the accretion rate capped at ten times the reference Eddington rate. The BH accretion disks are disrupted at each BH stochastic kick and the model accounts for both thin and slim disks depending on the accretion rate.
	
		\begin{figure}[tbp]
		\centering
		\includegraphics[width=0.8\textwidth]{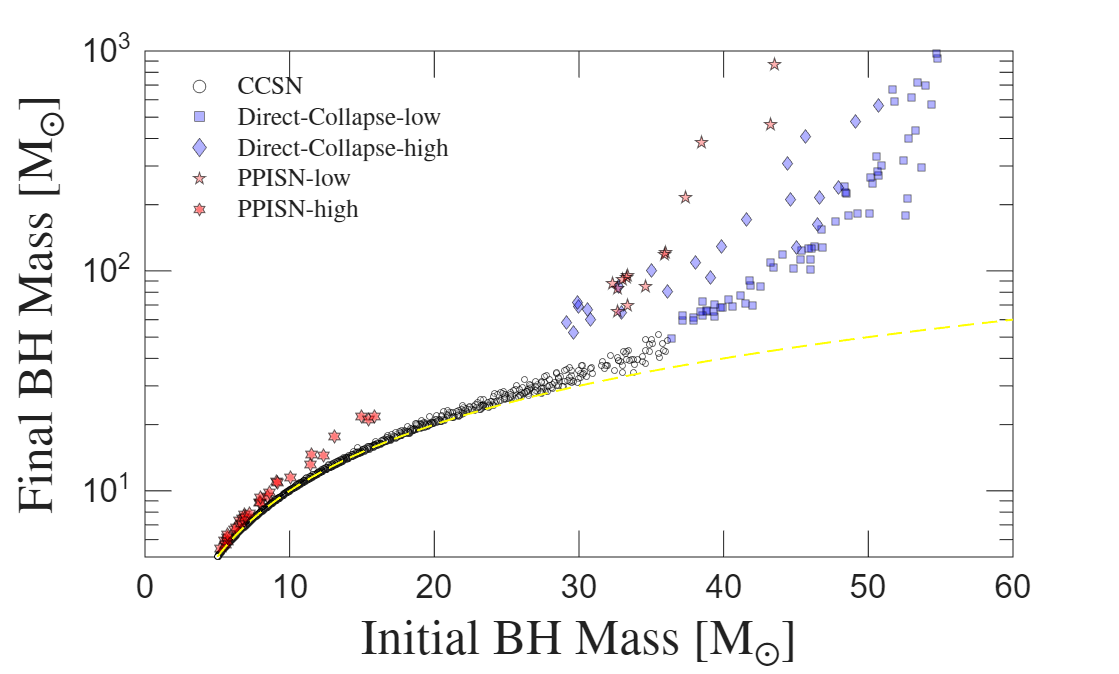}
		\caption{Mass scatter diagram for cluster stellar mass $M_{\star} = 10^6\,{\rm M}_\odot$, size $r_{c,\star} = 1 \,{\rm pc}$, star formation efficiency $\varepsilon = 0.35$, and one simulation run.}
		\label{fig:m_fin-ini}
	\end{figure}
	
		\begin{figure}
		\centering
		\includegraphics[width=0.8\textwidth]{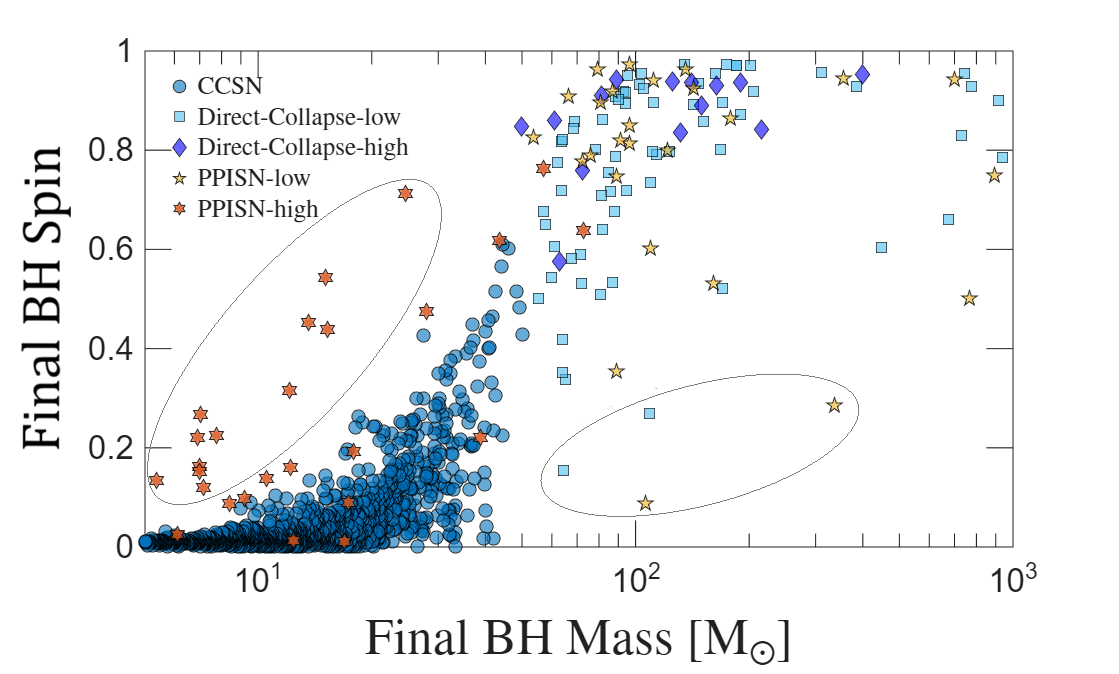}
		\caption{Final BH spin with respect to final BH mass of all BHs for one representative simulation run of our typical cluster with $M_{\star} = 10^6\,{\rm M}_\odot$, $r_{c,\star} = 1 \,{\rm pc}$. We have circled low-mass/intermediate-spin and high-mass/low-spin outliers. The legend labels provide the BH origin; CCSN for a core-collapse-supernova remnant; Direct-Collapse-low(high) for a BH generated by direct collapse of the core of a low(high) mass star; PPISN-low(high) for a BH remnant from a pulsational-pair-instability supernova of low(high) mass.}
		\label{fig:a_m_fin_data}
	\end{figure}
	
	Throughout this work, $\chi \equiv c S/(G m^2) \in [0,1]$ denotes the dimensionless spin magnitude of a BH of mass $m$ and spin angular momentum $S$.
	We adopted initial BH spins of $\chi_{\rm ini}=0.01$, consistent with the recent models of efficient angular momentum transport during core contraction of the massive star progenitor (revised Tayler instability)\cite{2019ApJ...881L...1F,Fuller_2019,2020A&A...636A.104B}. 
	Even in the case of natal spin $0.1$, consistent with the classic Tayler-Spruit dynamo\cite{Spruit_2002,Heger_2005,Qin_2018}, 
	the general form of the spin-mass relation persists with $\chi$ only shifted to values $\sim 0.1$ or lower for $m\lesssim 20\,{\rm M}_{\odot}$. The maximum spin, and the transition region remain nearly identical to our adopted $\chi_{\rm ini} = 0.01$ case. Our simulations give in effect thin to slim disks with dimensionless accretion rate (BH accretion rate over Eddington rate, see Ref.~[\refcite{2026A&A...709A.5R}]) $\dot{m} \sim 0.1-10$ depending on the gas density, along with the location and velocity of the BH. 
	We consider star clusters with masses $M_{\star} = (0.35-1) \cdot 10^{6}\,{\rm M}_\odot$, size $r_{c,\star} = 0.5-1.2\,{\rm pc}$, two metallicity values $Z = \{0.01,0.1\}\,Z_{\odot}$, and star formation efficiency $0.35$, which can generate at least one BH with mass $\sim 150\,{\rm M}_{\odot}$ covering the GW231123 primary component mass\cite{2025ApJ...993L..25A}. 
	
	The lighter BHs are more sensitive to gravitational perturbations, while the more massive BHs can form more persistent disks and accumulate closer to the center of the cluster where the gas density is higher. As a result, we identify a saturating spin--mass correlation with low spins at low masses, rising continuously with mass through all intermediate spin values, to high spins at high masses. In Fig.~\ref{fig:a_m_fin_data} we depict the final BH spins vs masses for a typical simulation run.
	The spin-mass relation is persistent for sufficiently massive and dense clusters with
	a median spin which can be consistently fitted with a saturated exponential law of the form
	\begin{equation}\label{eq:med_fit}
		\bar{\chi}_{\rm acc}(m) = \chi_{\rm max}\left(1 - e^{-\left(\frac{m}{m_{\rm T}}\right)^{\beta}}\right), 
	\end{equation}
	where the parameters vary only slightly. 
	For the various star-cluster configurations we considered, the maximum spin takes values $\chi_{\rm max} = 0.74-0.90$, the characteristic mass scale $m_{\rm T} = 40-55\,{\rm M}_{\odot}$, and the exponent $\beta = 2.5-3.3$.
	For the pooled sample of clusters we get the fit $\{\chi_{\rm max}, m_{\rm T}, \beta\} = \{0.83, 46.5\,{\rm M}_{\odot}, 2.85\}$, displayed as a dashed line in Fig.~\ref{fig:spinmass_acc_scatter}. 
	
	\section{Evidence in GW data}\label{sec:gwtc}
	
	\begin{figure}[ht]
		\centerline{\includegraphics[width=0.8\textwidth]{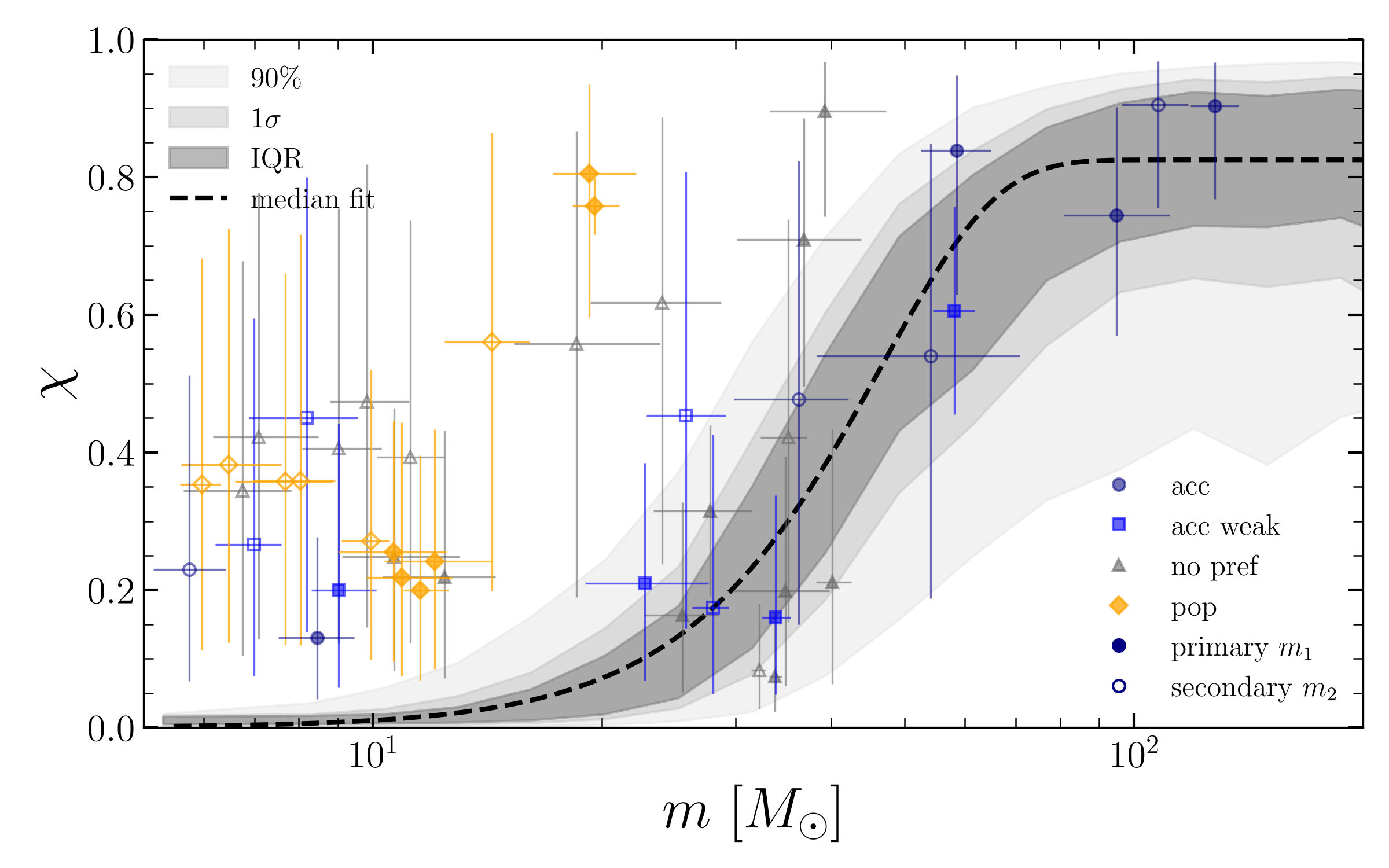}}
		\caption{Component spins $\chi$ of LVK binary black holes in the cumulative GWTC-5.0---restricted to spin-informative events---with respect to their mass $m$,
			against the spin--mass relation predicted by our accretion channel.
			Points: LVK posterior medians with $90\%$ credible intervals, filled markers for
			the primary and open for the secondary; classes:  `acc' for $\ln B^{\rm Def}>1$,  `acc weak' for natural log Bayes factor $0.5<\ln B^{\rm Def}\le1$, `no pref' for $|\ln B^{\rm Def}|\le0.5$, and `pop' for $\ln B^{\rm Def}<-0.5$. Bands and dashed curve: percentiles and
			median fit of the predicted $p_{\rm acc}(\chi\mid m)$ by simulations.}
		\label{fig:spinmass_acc_scatter}
	\end{figure}

	We investigated the cumulative GWTC-5.0 released by LVK\cite{2026arXiv260527225T} for Bayesian evidence of the accretion-driven spin-mass correlation\cite{2026arXiv260829072R}.
	We compared our prediction with three reference models. 
	Our null is the LVK \textsc{Default-bbh} model\cite{2026arXiv260527226T}, the fiducial description of the whole BBH population over GWTC-5.0, which draws both spins from a common truncated Gaussian, and carries no mass dependence. 
	The second comparative model, we call \textsc{2br}, is also a GWTC-5.0 population-informed model, which already asserts a rapidly spinning subpopulation.
	It is the spin sector of the Three Subpopulation model of Ref.~[\refcite{2026PhRvL.137b1403B}], which LVK refit to GWTC-5.0 and use to constrain the onset of a high-mass subpopulation\cite{2026arXiv260527226T} to
	$m^t_{\rm 2br} = 46.6^{+11.6}_{-6.7}\,{\rm M}_{\odot}$. The spin sector has two branches and the Gaussian switches between them at $m^t_{\rm 2br}$.
	The third comparative model, denoted \textsc{Merg}, is designed to capture the characteristic spin distribution of the hierarchical-merger channel.
	We assume that a component above the second-generation mass floor, taken at $9.5\,{\rm M}_{\odot}$, the lightest possible merger remnant, is itself a remnant with a marginalized probability, in which case its spin is drawn from the $\mathrm{Beta}(6.6,3.4)$ distribution of mode
	$0.7$ and concentration $\nu=10$, reproducing the near-universal remnant spin of a comparable-mass merger
	\cite{2008ApJ...684..822B,2017ApJ...840L..24F}; otherwise it follows \textsc{Default-bbh}.
	
	In Fig.~\ref{fig:spinmass_acc_scatter} we depict the predicted accretion-driven spin-mass distribution overlaid with the GW events of GWTC-5.0 with at least one BH component with informative spin $\Delta_{90}\chi_i\leq 0.63$. Over the whole GWTC-5.0 LVK identified $259$ GW mergers under a conservative selection\cite{2026arXiv260527226T}. Of those $259$ events, $10$ show positive evidence $\ln B^{\rm Def}>1$ (`acc') for the accretion-driven channel and a further $29$ reach $0.5<\ln B^{\rm Def}\le1$ (`acc weak'), spanning the full component-mass range of the catalogue. We further calculate the population mixture Bayes factor $\ln K^X$ against each model $X$ over several mass ranges. The evidence in favour of an accretion-driven subpopulation for several primary-mass bands is reported in Table~\ref{tab:massspan}.
	The LVK analysis\cite{2026arXiv260527226T} in the GWTC-5.0 suggests that rapidly-spinning subpopulations exist in two separate bands, $10\,{\rm M}_{\odot} \lesssim m_1 \lesssim 20\,{\rm M}_\odot$ and $m_1 \gtrsim 45\,{\rm M}_{\odot}$. Therefore, we included those bands in our results. The upper band is notably consistent with the characteristic mass $m_{\rm T}$ of our predicted median fit Eq.~(\ref{eq:med_fit}). 
	The upper edge of the LVK low-mass rapidly-spinning population,
	$\sim 20\,{\rm M}_\odot$, lies within the credible region of our inferred change point $m^t = 20.7^{+11.6}_{-1.2} \,{\rm M}_{\odot}$.
	The evidence in favour of the accretion channel is strongly concentrated above this representative boundary, $m_1\ge m^t$, with $\ln K^{\rm Def}=16.6$. For more details and a complete table of Bayes factors per GW event see Ref.~[\refcite{2026arXiv260829072R}].
	
\begin{table}[ht]
	\tbl{Mixture evidence by primary-mass span.}
	{\begin{tabular}{@{}lrrrr@{}} \toprule
			Span $[{\rm M_\odot}]$ & $N$ & $\ln K^{\rm Def}$ & $\ln K^{\rm 2br}$ & $\ln K^{\rm Merg}$ \\ \colrule
			$m_1\ge46$ & $58$ & $13.57$ & $4.69$ & $12.19$ \\
			$20\le m_1<46$ & $136$ & $2.80$ & $1.80$ & $4.15$ \\
			$10\le m_1<20$ & $59$ & $-2.13$ & $-2.29$ & $-2.01$ \\
			$m_1<10$ & $6$ & $3.77$ & $3.48$ & $3.80$ \\ \colrule
			all masses & $259$ & $8.76$ & $1.22$ & $8.99$ \\
			$m_1\ge20$ & $194$ & $16.59$ & $6.56$ & $16.84$ \\ \botrule
	\end{tabular}}
	\begin{tabnote}
		$N$: number of events in the span;
		$\ln K^{\rm Def}$, $\ln K^{\rm 2br}$, $\ln K^{\rm Merg}$: log mixture Bayes
		factors of the accretion model against the \textsc{Default-bbh}, \textsc{2br} models,
		and the hierarchical-merger model \textsc{Merg}.
	\end{tabnote}
	\label{tab:massspan}
\end{table}
	
	\section{Conclusion}
	
	Our analysis suggests that a certain spin-mass correlation of stellar BHs is generated as an integral element of the formation stage of dense massive star proto-clusters. The BHs accrete the residual gas from the first star-formation event and they have the time to grow in mass before the gas is depleted by stellar feedback for sufficiently compact clusters. An accretion disk around each BH is formed by the BH's own orbital motion as gas angular momentum, induced by the velocity shear between inner and outer hemispheres of the Bondi sphere, is injected into the Bondi sphere. The accretion disk can spin up the BH depending on its mass and general motion. The smoking gun of this process is a certain spin-mass correlation, which can be expressed as a saturating exponential of the medians\cite{2026A&A...709A.5R}, Eq.~(\ref{eq:med_fit}). We investigated the GWTC-5.0 released by LVK and found strong evidence of an accretion-driven subpopulation\cite{2026arXiv260829072R} concentrated above a primary mass scale $m_1\gtrsim 20\,{\rm M}_{\odot}$.
	
	\section*{Acknowledgments}
	
	ZR is supported by the European Union's Horizon Europe Research and Innovation Programme under the Marie Sk\l{}odowska-Curie grant agreement No.~101149270--ProtoBH. Numerical calculations have been made possible through a CINECA-INFN agreement, providing access to resources on Galileo100/LEONARDO at CINECA.
	
	\bibliographystyle{unsrt}
	\bibliography{ICNFP2026_protoBH}
	
\end{document}